      [1999/12/01 v1.4c Il Nuovo Cimento]
\documentclass{cimento}

\usepackage{amssymb}
\usepackage{natbib}
\def\araa{ARA\&A}%
\def\apj{ApJ}%
\def\apjl{ApJ}%
\def\aap{A\&A}%
\def\mnras{MNRAS}%
\def\nat{Nature}%
\usepackage{graphicx}  
\title{GRB progenitors at low metallicities}
\author{R.~Hirschi\from{ins:bas}\ETC,
G.~Meynet\from{ins:gva},
A.~Maeder\from{ins:gva}}
\instlist{\inst{ins:bas} Physics and Astronomy, University of Basel, Klingelberstr. 82, 4056 Basel, Switzerland
  \inst{ins:gva} Geneva Observatory, Ch. des Maillettes 51, 1290 Sauverny, Switzerland}
\PACSes{\PACSit{00.00}{By the way, which PACS is it, the 00.00? GOK.}
\PACSit{---.---}{\ldots}}
\begin{document}

\maketitle

\begin{abstract}
We calculated pre--supernova evolution models of single rotating massive stars. 
These models reproduce observations during the early stages of
the evolution very well, in particular Wolf--Rayet (WR) populations and 
ratio between type II and type Ib,c supernovae at
different metallicities ($Z$). Using these models we found the following results 
concerning long and soft gamma--ray burst (GRB) progenitors:
\begin{list}{-}{}
\item GRBs coming from WO--type (SNIc) WR stars are only produced at low $Z$ ($Z_{\rm LMC}$ or lower).
\item The upper metallicity limit for GRBs is reduced to 
$Z \sim 0.004$ (SMC) when the effects of magnetic fields are included.
\item GRBs are predicted from the second (and probably the first) stellar generation onwards.
\end{list}
\end{abstract}

\section{Introduction}
Gamma--Ray bursts (GRBs) can be divided into 
two main types: a) short and hard and b) long and soft 
\cite[see][]{Pi04}.
The long soft GRBs have now been firmly connected to supernova
(SN) explosions of the type Ic 
\cite[GRB 980425--SN 98bw, GRB 030329--SN 03dh, GRB 031203--SN 03lw, GRB 060218--SN 06aj; see for example][]{MNM03,WB06}.
The leading model for GRB--SN explosions is the collapsar model 
\cite{W93, FW99}. 
In this model, the progenitor star must first lose its hydrogen 
rich envelope during its lifetime but retain enough angular momentum in its core. 
The star must also form a black hole when it dies. The large amount of angular 
momentum induces the formation of an accretion disk around the black hole. 
Then accretion from the disk onto the black hole powers bipolar jets. 
These jets can only pierce through the outer part of the star and be detected 
if the star is compact. This explain why the progenitor star has to lose its hydrogen 
rich envelope during its lifetime.
Additional constraints come from the fact that
close-by SN associated GRBs are found 
\cite{Fl03,St06,Fru06} 
in low metallicity ($Z$) environments, 
corresponding to the metallicity of the Magellanic Clouds ($Z \sim 0.004-0.008$).
Low metallicity environments are also favoured from the theoretical point of view 
because lower metallicity means weaker mass loss and therefore smaller angular 
momentum loss 
\cite{FW99}.

Several studies were devoted to finding which stars can
lead to GRBs 
\cite{PMN04, HWS05,IRT04,grb05,PLYH05,FH05,WH06,YLN06}. 
The key ingredients of the progenitor models in the context of GRBs are
the treatment of rotation, magnetic fields, binarity and metallicity.
In this work, we look at GRB progenitors coming from single stars with models, 
which include the treatment of differential rotation and its transport in the interior. 
We focus on the metallicity dependence of the GRB rates and discuss the 
impact of magnetic fields on our results.

\section{Computer model \& calculations}
The stellar evolution code used for this work is described in
 \cite{psn04}. 
Convective stability is determined 
by the Schwarzschild criterion. Overshooting is only considered for 
H-- and He--burning cores with an overshooting parameter, 
$\alpha_{\rm{over}}$, of 0.1 H$_{\rm{P}}$. 
The mass loss is proportional to the square root of the metallicity. 
It is also dependent on the surface rotation velocity.
The centrifugal force is included in the structure equations.
Last but not least, the processes taken into account, which induce transport and mixing of angular 
momentum and matter, are meridional circulation and dynamical and secular 
shears. Magnetic fields are included only in the models discussed in the 
section on magnetic fields. For these models, 
the treatment of magnetic fields 
\cite{Sp02} is included according to \cite{ROTBIII}. 
Models were calculated at various metallicities: 
$Z=0.04$ (GC), 0.02 ($Z_\odot$), 0.008 (LMC), 0.004 (SMC), and $10^{-8}$ 
(second stellar generation)
\cite[see][for more details]{grb05,H06}.

\section{Theoretical GRB rates around solar metallicities}
Our models predict the production of fast rotating BHs. 
Models with large initial masses or high metallicity end with less angular momentum
in their central remnant with respect to the break--up limit for the remnant.
Many WR star models satisfy the three main criteria (BH formation,
loss of hydrogen--rich envelope and enough angular momentum to form an
accretion disk around the black hole) for long and soft gamma--ray bursts (GRB) 
production via the collapsar
model 
\cite{W93}. Considering all types of WR stars as GRB progenitors, 
we predict too many GRBs compared to observation. 
If we consider only 
WO--type WR stars (type
Ic supernovae as is the case for the GRB related SNe discovered so far, 
see the list in the introduction) as
GRB progenitors, the GRB production rates are in much better agreement
with observations (see Table \ref{grbrates}). In particular, WO stars are found only at low
metallicities (Z$\lesssim$Z$_{\rm LMC}$) and have masses, 
$M \gtrsim$ 50 $M_\odot$.

\begin{table}
  \caption{Predicted rates [yr$^{-1}$] of 
long soft GRBs coming from WO--type WR stars,
as well as limiting masses at various metallicities.}
  \label{grbrates}
\begin{tabular}{l r r r r}
\hline
& $Z_{\rm SMC}$   &   
  $Z_{\rm LMC}$   &   
  $Z_{\odot}$     &   
  $Z_{\rm GC}$       
  \\
 \hline 
$M^{\rm min}_{\rm GRB}$(WO) & 50 & 45 & -  & - \\
$M^{\rm max}_{\rm GRB}$(WO) & 95 & 95 & -  & - \\
  ${\cal R}_{\rm GRB}$(WO) 
& 4.74E-04 
& 5.99E-04 
& 0  
& 0  \\
 \hline
\multispan{5}{\hfill Observed rate:
 ${\cal R}^{\rm OBS}_{\rm GRB} = 3\,10^{-6}-6\,10^{-4}$ (depending on the beaming angle of GRBs) \hfill } \\ \\
\multispan{5}{\hfill Reference total SN rate: 
${\cal R}^{\rm OBS}_{\rm SN}\simeq 7\,10^{-3}$ \cite{PMN04} \hfill } \\
 \hline \\
\end{tabular} 
\\
\end{table}

\section{Do the first stars produce GRBs?}
\begin{figure}
\includegraphics[width=7cm]{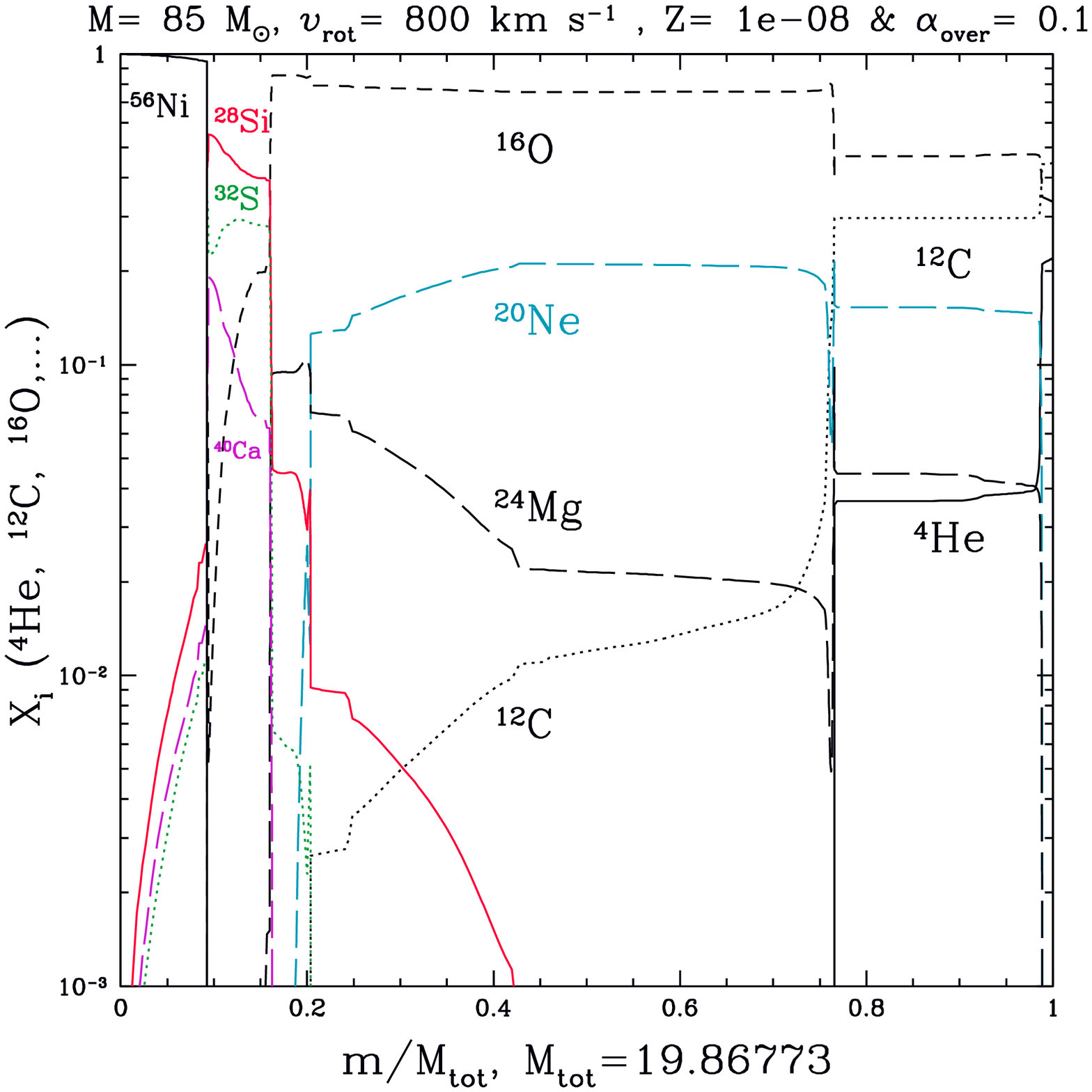}\includegraphics[width=7cm]{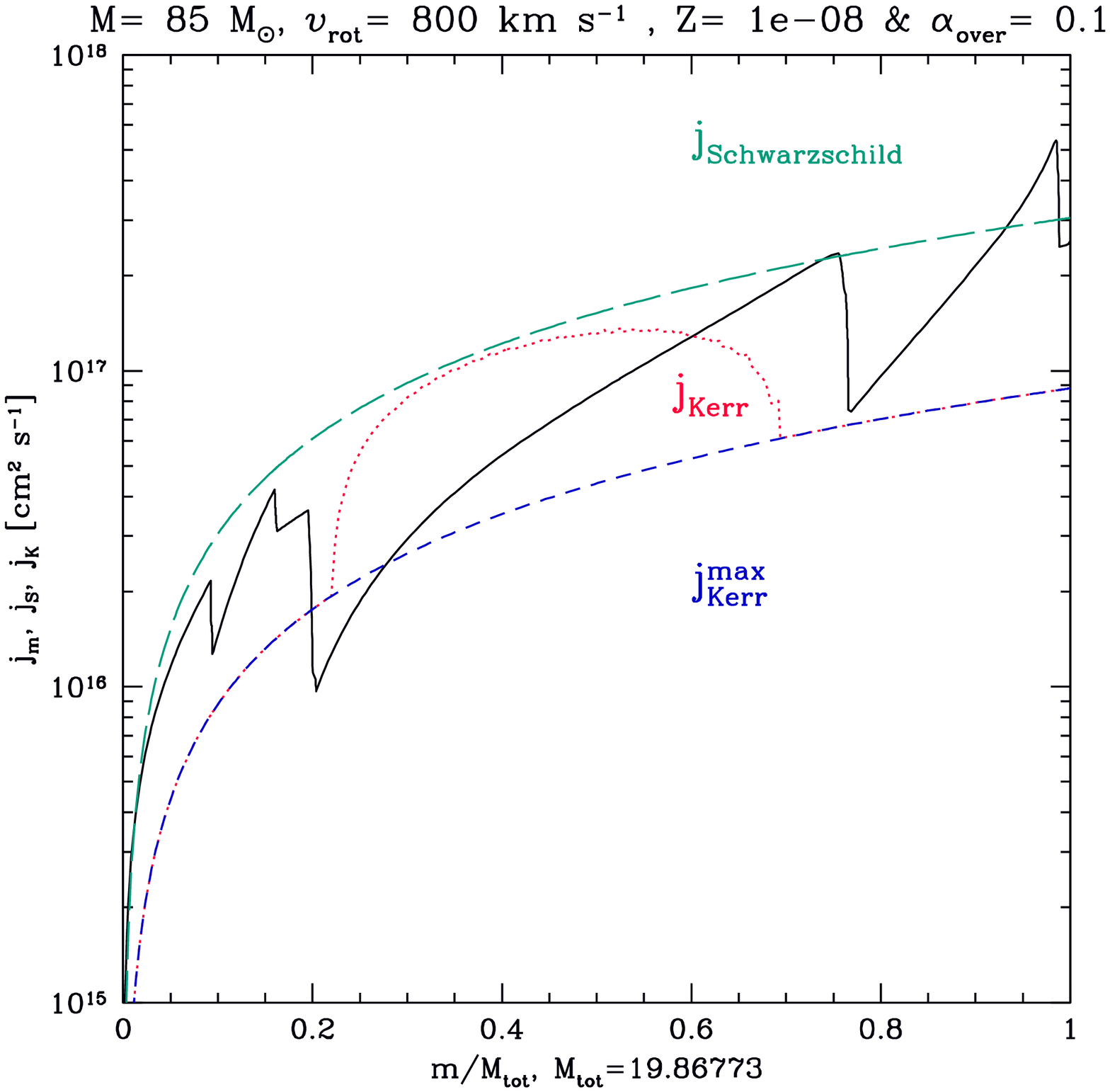}
\caption{
{\it Left}: profiles of the chemical abundances at the pre--SN stage
for the 85 $M_\odot$ model at $Z=10^{-8}$.
{\it Right}: profile of the specific angular momentum 
(solid line) for the same model. The red dotted line 
shows the minimum angular momentum necessary in
order to form an accretion disk around a rotating black hole.
The blue short dashed and green long dashed lines show the minimum angular momentum
necessary for matter around a maximally rotating and a non--rotating black hole,
respectively.}
\label{85maj}
\end{figure}

We calculated a series of models at $Z=10^{-8}$ 
(which corresponds to the second stellar generation).
In this series, the 85 $M_\odot$ model  
becomes a WO--type WR star (see Fig. \ref{85maj} {\it left}). 
The core of the 85 $M_\odot$ model retains enough
angular momentum to produce a GRB via the collapsar model (see Fig. \ref{85maj} {\it right}). 
SNe of type Ib,c and GRBs are
therefore predicted to ensue from the death of single massive stars at
very low metallicities. 

\section{Models with rotation and magnetic fields}

Petrovic et al \cite{PLYH05} and Fryer and Heger \cite{FH05}
studied the evolution of single and binary massive stars with and without magnetic fields 
\cite[dynamo theory from][]{Sp02}. 
They find that both single
and binary models without magnetic field can produce GRBs and both 
single and binary models with magnetic field considered in their studies 
cannot produce GRBs. Models with magnetic fields nevertheless better reproduce the observed
pulsar periods.

The most recent studies found a way to produce GRBs in models 
including magnetic fields. Indeed, \cite{YLN06} and \cite{WH06}
find that if a star initially rotates fast enough, it undergoes quasi chemical homogeneous
evolution. The key idea is that WR stars are produced by mixing and not by mass loss.
Since WR mass loss rates are too high at metallicities around solar, GRBs can only be produced
via this channel in low metallicity environments. \cite{YLN06} find that
GRBs can be produced for $Z \lesssim 0.004$. This is somewhat lower than the 
observed metallicity of close-by GRBs.

We calculated 40 $M_\odot$ models at $Z=Z_{\rm SMC}$ and $Z=10^{-8}$ with magnetic fields. 
At $Z=Z_{\rm SMC}$, we obtain similar results as \cite{YLN06}.
At $Z=10^{-8}$, we found that 40 $M_\odot$ models undergo quasi chemical homogeneous 
evolution (if the initial velocity is larger than 600\,km\,s$^{-1}$). 
With or without magnetic fields, GRBs are therefore expected to follow 
the death of massive stars from the second stellar generation onwards (and probably from the first).

The treatment of anisotropic mass loss and the recent downward revision of solar metallicity 
are two important physical ingredients that have yet to be included in the models and that will 
help GRB progenitors retain more angular momentum.


\appendix
\section{Questions and answers:}

\noindent{\it Question by A. J. van Marle}: In several GRB afterglows, silicon is seen in absorption at a velocity
that indicates its presence in the progenitor wind. Is this possible for low metallicity stars?
\\ {\it Answer}: Yes, pristine silicon is present in the stellar
wind. Nearby GRBs probably do not have very low metallicity progenitors, in which case pristine silicon
could be detected.
\\ 
\\{\it Question by D. Fox}: You have investigated models of isolated massive stars, but many massive stars
are born in close binaries. Does this affect your results?
\\ {\it Answer}: The presence of a binary companion may either increase mass loss or induce
accretion (at different times) and spin--up or spin--down a star. This makes it difficult to estimate the
final effect for GRBs and obtain quantitative predictions.
\\ 
\\{\it Question by C. Thompson}: Could you summarize how your model of angular momentum transport is
constrained by observations of isotope enhancements in evolved stars?
\\ {\it Answer}: One important test is comparing the surface enrichment, for example in nitrogen, between
models and observations as a function of the position in the Hertzsprung--Russell diagram.
\\ 
\\{\it Comment by T. Piran}: According to your low metallicity paradigm, we need new estimates of GRB
rates based on the SFR for low metallicity stars.
\\ {\it Answer}: I agree with this comment. Langer, Norman and collaborators have started to study this
point.

\end{document}